\documentclass[a4paper,12pt,notitlepage]{article}

\usepackage{amsmath}
\usepackage{latexsym}
\usepackage{amssymb}
\usepackage{amsfonts}
\usepackage{times}

\usepackage[dvips]{graphicx}

\newcommand{\R}{ \mathbb{R} }
\newcommand{\C}{ \mathbb{C} }
\newcommand{\CP}{ \mathbb{C}_{+} }
\newcommand{\re}{\mathop{\mathrm{Re}}}
\newcommand{\im}{\mathop{\mathrm{Im}}}
\newcommand{\tr}{\mathop{\mathrm{tr}}}

\newtheorem{Thm}{Theorem}[section]
\newtheorem{Lemma}[Thm]{Lemma}
\newtheorem{Prop}[Thm]{Proposition}

\author{Ilya Ya. Goldsheid and Boris A. Khoruzhenko\\
        School of Mathematical Sciences, Queen Mary,
        University of London,\\ London E1 4NS, U.K.}

\title{Regular Spacings of Complex Eigenvalues in the
One-dimensional non-Hermitian Anderson Model}
\date{23 August 2002}

\begin{document}

\maketitle
\begin{abstract}
We prove that in dimension one the non-real eigenvalues of the
non-Hermitian Anderson (NHA) model with a selfaveraging potential
are regularly spaced. The class of selfaveraging potentials which
we introduce in this paper is very wide and in particular includes
stationary potentials (with probability one) as well as all
quasi-periodic potentials. It should be emphasized that our
approach here is much simpler than the one we used before. It
allows us a) to investigate the above mentioned spacings, b) to
establish certain properties of the integrated density of states
of the Hermitian Anderson models with selfaveraging potentials,
and c) to obtain (as a by-product) much simpler proofs of our
previous results concerned with non-real eigenvalues of the NHA
model.
\end{abstract}

\section{Introduction}\label{sec1}
The non-Hermitian Anderson model (NHA model) was introduced by N.
Hatano and D. Nelson in 1996. It arises in the physics of vortex
matter, \cite{HN1,HN2}, and in many other contexts, see e.g.
\cite{E,NS,SN}. This model is described by the following operator
\begin{equation}\label{main}
(H_n^g\varphi)_k=-e^g\varphi_{k+1}+q_k\varphi_k-e^{-g}\varphi_{k-1},\
\ 1\leq k\leq n
\end{equation}
with periodic boundary conditions
\begin{equation}\label{bc}
\varphi_0=\varphi_n,\ \varphi_1=\varphi_{n+1}.
\end{equation}
Here $g$ is a real parameter, $g\geq 0$. The Hilbert space is
$l_2(1,n)$ with the standard inner product: if
$\varphi=\{\varphi_j\}_{j=1}^n$ and $\psi=\{\psi_j\}_{j=1}^n$ are
two vectors from $l_2(1,n)$, then
$(\varphi,\psi)=\sum_{j=1}^n\varphi_j\bar{\psi}_j$.

\begin{figure}[ht]

\label{fig1} \centerline{\includegraphics[width=14cm]{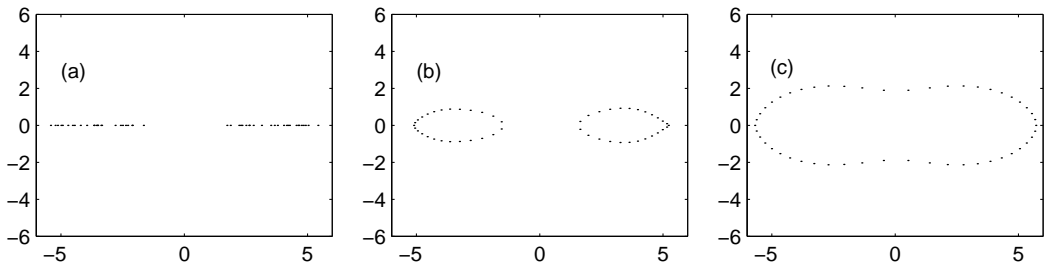}}
\caption{The eigenvalues of $H_{n}^g$, are presented by dots plotted 
in the complex plane. Here $n=50$ and $\{q_k\}$ is a fixed realization
 of independent samples from the uniform distribution on
$[-4,-3]\cup [3,4]$; the values of $g$ are: (a) $g=0.2$, (b)
$g=1.1$, (c) $g=1.4$.}
\end{figure}

Hatano and Nelson considered the case when the values $q_j$ of the 
potential are taken as a realization of a sequence of independent
identically distributed random variables. 
By conducting a numerical experiment they
discovered a number of remarkable properties both of the spectrum
and the eigenfunctions of the operator $H_n^g$. 
It turns out that the asymptotic behavior of the eigenvalues depends
strongly on the value of the parameter $g$. To demonstrate this statement
we present in Fig.\ref{fig1} results of a similar numerical experiment.
These pictures are not so well predictable in the following sense.
It is a consequence of the Weyl criterion that the spectrum 
of the limiting random operator ($n=\infty$) contains with 
probability 1 the union of spectra of operators with constant  
potentials, $q_j\equiv q$, for any real $q$ belonging  to the
support of the random variable $q_1$. A very simple calculation shows then 
that the spectrum of $H_{\infty}^g$ would typically contain 
(with probability 1) a two-dimensional subset of the complex plain.
(A much more detailed description of the spectrum of the limiting 
operator can be found in \cite{D}.)  
However, numerical experiments reproduce pictures like that 
in Fig.\ref{fig1} with remarkable stability also for large values 
of $n$ (in \cite{HN1,HN2}
$n=1000$). They clearly show that the eigenvalues of $H_{n}^g$ have 
no tendency to spread  over any two-dimensional region but rather tend 
to belong to smooth curves.

Our attempt to understand whether this phenomenon would indeed persist
as $n\to\infty$ or is due to the fact that
$n$ is not large enough stimulated the appearance of two 
papers \cite{GK1,GK2} where the 
analysis of the spectra of operators $H_{n}^g$ for finite but large 
values of $n$ was carried out. We shall now briefly describe some of
our results restricting ourselves to the case of bounded potentials. 
Namely it turns out that there are critical values
$\underline{\hspace{0.5ex}g}_{\mathrm{cr}}$ and $\bar{g}_{cr}$ depending 
only on the distribution of the random potential and such that:

1. If $0\leq g<\underline{\hspace{0.5ex}g}_{\mathrm{cr}}$ then 
the eigenvalues of $H_n^g$ are ``asymptotically'' real 
(see Theorem \ref{sec2:thm0} for the exact meaning of this statement).
Moreover their limiting distribution does not depend on $g$ and is
the same as in the Hermitian case (that is with $g=0$).

2. If $\underline{\hspace{0.5ex}g}_{\mathrm{cr}}<g<\bar{g}_{cr}$ 
then a finite proportion of
eigenvalues moves out of the real axes and places itself on very
smooth curves in the complex plane. These curves converge 
to non-random limiting curves as $n\to\infty$.
Moreover we found the
limiting density of states for non-real eigenvalues and proved that
the ``asymptotically'' real eigenvalues have the same limiting 
distribution as in the case of the self-adjoint model.

3. If $g>\bar{g}_{cr}$ then virtually all eigenvalues 
leave the real axes.

One thus concludes that the phenomenon described above persists 
as $n$ is growing. The fact that the spectrum of the limiting operator 
is two-dimensional means that on this spectrum (but off the 
above mentioned curves) the resolvent of $H_n^g$ exists but its
norm tends to infinity as $n\to\infty$. (See also \cite{TCE} 
where the spectrum and the behavior of the norm of the 
resolvent of random bidiagonal matrices were studied.)

In the present paper we investigate spacings between the
neighboring non-real eigenvalues. We do this for models with
selfaveraging ({\it deterministic}!) potentials which are introduced below. 
The class of selfaveraging potentials is very wide and includes 
in particular random stationary, quasi-periodic, and many other potentials.

The approach used in \cite{GK1,GK2} was based on the theory of
products of random matrices and on the potential theory. We would
like to emphasize that the present paper is self-contained and
that the technique we use here is much simpler and more
straightforward than that we used before. It allows us:

- To prove that the non-real eigenvalues of NHA model behave in a
very regularly way: not only do they belong to very smooth curves as 
$n\to\infty$ but also the difference between any two neighboring 
non-real eigenvalues $z_{k+1}$ and $z_k$ can be calculated 
with a great deal of precision as 
\[
z_{k+1}-z_k=\frac{2\pi i}{f(z_k)}\frac{1}{n},
\]
where $f(z)$ is an analytic function of $z$ (Theorem \ref{sec2:thm4}). 
This is our principal new result.

- To establish the log-H\"older continuity of the density of
states for Hermitian Anderson models with selfaveraging
potentials.

- To obtain (essentially as a by-product) much simpler proofs of
our previous results listed above.

\emph{Remarks.}
1.  In \cite{GK1,GK2} we considered tri-diagonal
matrices with off-diagonal elements $H(j,j+1)$ and $H(j,j-1)$ depending
on $j$. All results of this paper can be extended to these models. 
The main additional condition that is needed 
is the existence of finite limits
$\lim_{n\to\infty}n^{-1}\sum_{j=1}^{n}\ln H(j,j+1)$ and
$\lim_{n\to\infty}n^{-1}\sum_{j=2}^{n}\ln H(j,j-1)$.

2. We have already mentioned above that, apart of the fact that
stationary potentials provide natural examples of
self-averaging potentials, randomness does not play any role in this paper. 
However as soon as one wishes to understand the real meaning
of the next order term in formula (\ref{z})
randomness becomes crucial. The very same thing applies to the properties of
asymptotically real eigenvalues. Namely, it is natural to conjecture 
that the asymptotically real eigenvalue in fact are real for sufficiently 
large values of $n$. However there are reasons to believe that in order 
to prove this conjecture one should restrict himself to the class of 
random potentials with good properties.

3. The approach based on the theory of products of random matrices 
(TPRM) is more difficult than that of the present work. However, 
all main results of this paper can be obtain within the framework of 
the TPRM. Moreover this is what we did first.
The additional advantage of the TPRM approach is that 
the case of one-dimensional differential Schroedinger operators
can be treated by this method in exactly the same way as 
the discrete case. The attempt to extend the approach of this
paper to the case of continuous space would lead to the necessity of
using a certain regularization procedure similar to the one used in \cite{CS}.

The paper is organized as follows. We introduce the class of
selfaveraging potentials and discuss the log-H{\"o}lder continuity
of the relevant density of states in Section 2. Section 3 contains
the statement of main results which are proved in Section 4.
In Appendix A we prove a version of a well known property of 
Lyapunov exponents (which we use in Section 2).
This allows us to (a) make our exposition even more self-contained
and (b) to demonstrate one more application of the technique we use.
Appendix B contains an example of calculation of Lyapunov exponents for 
models whose potentials have rare high peaks. We believe that these
 examples are of their own interest but the initial purpose of 
finding them was to provide a natural completion for the study of 
selfaveraging potentials.

\section{The selfaveraging potentials} \label{sec2}

We introduce now a class of \emph{deterministic} potentials for
which the distribution of the eigenvalues of the NHA model
(\ref{main}) -- (\ref{bc}) will be investigated in this paper.

Given an infinite sequence of real numbers $q\equiv
\{q_k\}_{k=1}^{\infty}$, we consider the sequence of selfadjoint
operators $H_n^0$, $n=2,3,\ldots$, with potential $q$. Let $N_n(E)$
be the distribution function of the eigenvalues of $H_n^0$,
\begin{equation}\label{sec1:2}
N_n(E)=\frac{1}{n}\#\{E_i\ :\ E_i\in\hbox{spectrum of $H_n^0$ and
$E_i<E$}\}.
\end{equation}

\medskip

\noindent {\bf Definition} \emph{ We say that a real potential $q$
is selfaveraging if the integrated density of states of the
self-adjoint Anderson model with this potential exists, i.e. there
exists a non-decreasing function $N(E)$ such that
$N_n(E)\longrightarrow N(E)$ as $n\to\infty$ at the points of
continuity of $N(E)$.}

\medskip

\emph{Remarks:} 1. We have borrowed the name `selfaveraging' from
the theory of random operators, where it is normally associated
with convergence of $N_n(E)$ to a nonrandom limiting function.

2. The class of selfaveraging potentials is very wide. For
example, it contains decaying potentials, periodic and almost
periodic potentials, and stationary random potentials\footnote{If
$\{q_k\}_{k=1}^{\infty}$ is a strictly stationary sequence of
random variables then $N_n(E)$ is weakly converging for almost all
realizations of $q$ with respect to the corresponding probability
measure.}, see e.g. book \cite{PF} for proofs and more examples.

3. We do not require $\int dN(E)=1$. However, in many cases
(and in particular in those mentioned above) the sequence of
measures $dN_n(E)$ is tight and hence cannot lose mass.

4. The choice of periodic boundary
conditions for $H_n^0$ is not essential, for example
one could use the Dirichlet boundary conditions
$\varphi_0=\varphi_{n+1}=0$ instead. This is because changing
boundary conditions amounts to a rank two perturbation of $H_n^0$,
and hence has no effect on $N(E)$ provided the perturbed finite
interval operators remain selfadjoint. Our preference for the
periodic boundary conditions will become apparent in the next
section.

\smallskip


Let
\begin{equation}\label{2a}
U_n(x,y)\overset{\rm def}{=}\int_{-\infty}^{+\infty} \ln |x+iy-E|
dN_n(E) \equiv \frac{1}{n}\sum_{j=1}^n \ln |x+iy-E_j|,
\end{equation}
where the summation is over all eigenvalues of $H_n^0$, and
\begin{equation}\label{1a}
U(x,y)\overset{\rm def}{=}\int_{-\infty}^{+\infty} \ln |x+iy-E|
dN(E),  \hspace{3ex} x,y\in \R,
\end{equation}
Obviously the logarithmic integrals $U(x,y)$ and $U_n(x,y)$ are
the real parts of the analytic functions
\begin{equation}\label{3}
F(z)\overset{\rm def}{=}\int_{-\infty}^{+\infty} \ln (E-z)
dN(E)=U(x,y)+iV(x,y).
\end{equation}
\begin{equation}\label{2}
F_n(z)\overset{\rm def}{=}\int_{-\infty}^{+\infty} \ln (E-z)
dN_n(E)=U_n(x,y)+iV_n(x,y).
\end{equation}
Here and below we consider analytic functions defined in the upper
half-plane $\CP=\{z\in \C:\hspace{1ex}  \im z>0 \}$,
and the branch of the $\ln(E-z)$ is chosen so that
$\ln (-i)=-i\frac{\pi}{2}$.

The functions defined in (\ref{2a}) -- (\ref{2}) play an important
role in this paper. In this section we study their properties
under the following conditions
\begin{itemize}
\item[{\bf C1}] \hspace{2ex} The potential $q$ is selfaveraging.
\item[{\bf C2}] \hspace{2ex} $\sup_{n\geq 1}\frac{1}{n}
\sum_{k=1}^n\ln(1+|q_k|) \le C < +\infty$.
\end{itemize}
The main role of Condition C2 is to ensure that the functions
$U(x,y)$ and  $F(z)$ are well defined:

\medskip

\begin{Prop} Assume C1-C2. Then for every $x,y\in\R$
the integral in (\ref{1a}) is converging.
\end{Prop}
\emph{Proof.} Suppose first that $y>0$ and let $z=x+iy$. Note that
\begin{equation}\label{i7}
\ln y \le U_n(x,y)\le \ln (2+|z|)+C \hspace{3ex} \hbox{for all
$x\in \R$ and $y>0$.}
\end{equation}
The LHS inequality is trivial, and the RHS inequality is ensured
by Condition C2. Indeed, 
\[
U_n(x,y) \le \frac{1}{n}\sum_{j=1}^n \ln (2+|z|+|q_j|),
\]
where the last inequality follows, e.g., from the representation
$U_n(x,y) =\frac{1}{n} \ln|\det(zI_n-H_n^0)|$ and Hadamard's
inequality for the determinants. By Condition C2,
\[
\frac{1}{n}\sum_{j=1}^n \ln (2+|z|+|q_j|) \le \ln (2+|z|) +C.
\]

If $A>-\infty$ and $B<+\infty$ are points of continuity of $N(E)$
then, in view of Condition C1,
\begin{eqnarray*} \int_{A}^{B} \ln |z-E|dN
(E)=\lim_{n\to\infty} \int_{A}^{B} \ln |z-E|dN_n(E)  &\le&
\lim_{n\to\infty} \int_{A}^{B} \ln |z+i-E|dN_n(E) \\ &\le &
\limsup_{n\to\infty} U_n(x,y+1).
\end{eqnarray*}
Applying (\ref{i7}), we obtain that
\begin{equation}\label{bB}
\int_{A}^{B} \ln |z-E|dN (E) \le \ln (3+|z|) +C.
\end{equation}
As $y\not=0$, the function $\ln |z-E|$ is bounded from below,
and (\ref{bB}) implies that the integral in
(\ref{1a}) is converging for $y>0$. By the
symmetry, it is also converging for $y<0$.

We shall now make use of the following inequality which will be
proved later (Theorem \ref{sec1:thm1}): $U(x,y)\ge -c_0$ for some
$c_0>0$ and all $x$ and $y\not=0$. This inequality together with
the monotone convergence theorem yield that
\begin{equation}\label{i4}
\lim_{y\downarrow 0} U(x,y) =\int_{-\infty}^{+\infty}\ln|x-E|dN(E)
\ge -c_0,
\end{equation}
hence the integral in (\ref{1a}) is also converging for $y=0$.
\hfill $\Box$

\medskip

\emph{Remark.} It follows from (\ref{i4}) that $N(\cdot)$ is a
continuous function. Therefore, under Conditions C1
and C2, we have that $N_n(E)$ converges pointwise to $N(E)$ as
$n\to\infty$, and the convergence is uniform in $E,\
-\infty<E<\infty$.

\smallskip

In view of the inequalities
\begin{equation}\label{ll}
Q_n-2 \le H_n^0 \le Q_n+2
\end{equation}
where $Q_n$ is the operator of multiplication by $q$,
$(Q_n\varphi)_j=q_j\varphi_j$, $j=1, \ldots , n$, Condition C2
also ensures that the sequence of measures $dN_n(E)$ is tight:

\medskip

\begin{Prop} Suppose that $q$ satisfies Condition C2. Then for
any $\varepsilon >0$ there exists $B>0$ such that $N_n(B)-N_n(-B)>
1-\varepsilon$ for all $n$.
\end{Prop}
\emph{Proof.} Denote by $\chi_{B}(E)$ the indicator-function of
the interval $[B, +\infty)$, and let $E_1, \ldots, E_n$ be the
eigenvalues of $H_n^0$. Then for any $B>0$
\[
1-N_n(B)=\frac{1}{n}\sum_{j=1}^n \chi_{B}(E_j) \le
\frac{1}{n}\sum_{j=1}^n \chi_{B}(E_j)\frac{\ln (1+E_j)}{\ln
(1+B)}\le \frac{1}{n}\sum_{j=1}^n  \chi_{B}(2+q_j) \frac{\ln
(1+2+q_j)}{\ln (1+B)}
\]
where the last inequality follows from (\ref{ll}). We use here the
following fact: if $f(E)$ is a non-decreasing function and $A\le
B$ then $\tr f(A) \le \tr f(B)$. Since $\chi_{B}(2+q_j) \ln
(1+2+q_j)\le \ln (1+|2+q_j|)$ we conclude, in view of Condition
C2, that
$
N_n(B) \ge 1- C^{'}/\ln (1+B)
$
for some constant $C^{'}>0$ and all $n$. Similarly, $N_n(-B)\le
C^{''}/\ln (1+B)$ for some constant $C^{''}>0$ and all $n$. \hfill
$\Box$

\medskip

We shall now investigate the relation between $F_n(z)$ and $F(z)$
in the limit $n\to\infty$.

\medskip

\begin{Prop} \label{prop:liminf} Assume C1-C2. Then for every real $x$ and $y\not=0$
\begin{equation}\label{l}
\liminf_{n\to\infty}U_n(x,y) \ge U(x,y).
\end{equation}
\end{Prop}
\emph{Proof.} For all $B$ large enough (so that
$[x-1,x+1]\subseteq [-B,B] $) we have that for any $n$
\[
U_n(x,y) \geq \int_{-B}^B \ln |x+iy -E| dN_n(E)
\]
and, by Condition C1,
\[ \liminf_{n\to\infty} U_n(x,y) \ge \int_{-B}^B \ln |x+iy -E|
dN (E).
\]
By letting $B\to\infty$, we obtain (\ref{l}). \hfill $\Box$

\medskip

Under Conditions C1 and C2, the sequence $\{U_n(x,y)\}$ is not
necessarily converging, even for $y\not=0$, and for some
selfaveraging potentials the inequality in (\ref{l}) is strict,
for examples see Appendix \ref{appendix2}.

In view of (\ref{i7}), for any compact set $K\subset \CP$,
$\sup_K|F_n(z)|\le M(K)$ for some constant $M(K)<+\infty$ and all
$n$. Hence the sequence $\{F_n(z)\}$ has a uniformly converging
subsequence. We shall now describe all limit points of
$\{F_n(z)\}$:

\medskip

\begin{Thm} \label{sec1:thm2} Assume C1-C2. Suppose that
$F_{n_j}(z)$ is a converging subsequence of $\{F_n(z)\}$. Then
necessarily
\begin{equation}\label{fc}
\lim_{j\to\infty} F_{n_j}(z)=F(z) + c
\end{equation}
for all $z\in \CP$ and some real constant $c$ satisfying the
inequality $0\le c \le \ln 3 +C$, where $C$ is the constant defined
in Condition C2.
\end{Thm}
\emph{Proof.} We note that $F^{'}_{n}(z)=\int_{-\infty}^{+\infty}
(z-E)^{-1}dN_n(E)$ and $F^{'}(z)=\int_{-\infty}^{+\infty}
(z-E)^{-1}dN(E)$. Since the function $h(E)=(z-E)^{-1}$ decays to
zero at infinity, Condition C1 ensures that
\[
\lim_{n\to\infty} F^{'}_{n}(z)=F^{'}(z)
\]
uniformly in $z$ on compact sets in $\CP$. Therefore there exists
a sequence of complex constants $c_n$ such that $F_n(z)-c_n
\longrightarrow F(z)$ as $n\to\infty$ for all $z\in \CP$. Passing
on to the converging subsequence $F_{n_j}(z)$, we have that
$c_{n_j}$ is also converging. Putting $c=\lim_j c_{n_j}$ we arrive
at (\ref{fc}). It remains to prove that $c$ is real, non-negative
and satisfies the inequality $c\le \ln 3 +C$.

Due to our choice of the branch of the log-function we have that
$-\pi \le \im \ln (E-z) \le 0$ for all $z\in \CP$. Thus
\[
\im F_n(z)=\int \im \ln  (E-z) dN_n(E)\longrightarrow \im F(z),
\hspace{3ex} \hbox{as $n\to\infty$},
\]
because the integrand is bounded, the sequence of measures
$dN_n(E)$ is tight and we have Condition C1. Therefore the
constant $c$ is real. The fact that it is non-negative follows
from Proposition \ref{prop:liminf}. To complete the proof, note
that $U_{n_j}(x,y)$ converges to $U(x,y)+c$ in the upper half of
the $xy$-plane. Therefore, because of (\ref{i7}), $U(x,y)+c \le
\ln (2+|z|)+C$. Putting here $x=0$ and $y=1$, we obtain $c\le \ln
3 +c -U(0,1)\le \ln 3 +C$. (Obviously, $U(x,1)\ge 0$ for all $x$.)
\hfill $\Box$

\medskip
The following condition
\begin{itemize}
\item[{\bf C2*}] \hspace{2ex} for any $\varepsilon >0$ there
is a $B>0$ such that
$
\frac{1}{n}\sum\limits_{j=1}^n \chi_B(|q_j|)\ln (1+|q_j|) <
\varepsilon
$
for all $n$
\end{itemize}
guarantees the convergence of $F_n(z)$ to $F(z)$. It is obvious
that Condition C2*  is somewhat more restrictive than C2. On the
other hand it is satisfied by many popular classes of potentials.
For example, the random stationary potentials with finite
expectation of $\ln (1+|q_j)$ satisfy Condition C2* with
probability one. It is also satisfied if
\[
\frac{1}{n}\sum_{j=1}^n \ln^{1+\delta}(1+|q_j|) \le C^{*} <
+\infty \hspace{3ex} \hbox{for some $\delta > 0$.}
\]

\medskip

\begin{Prop}\label{conv} Assume C1 and C2*. Then
\begin{equation}\label{conv1}
F_n(z)\underset{n\to\infty}{\longrightarrow} F(z) \hspace{2ex}
\hbox{uniformly in $z$ on compact sets in $\CP$}
\end{equation}
and, in particular,
\begin{equation}\label{conv2}
U_n(x,y)\underset{n\to\infty}{\longrightarrow} U(x,y) \hspace{2ex}
\hbox{uniformly in $z=x+iy$ on compact sets in $\CP$}.
\end{equation}
\end{Prop}
\emph{Proof.} If the potential $q$ is bounded then the statement
of Proposition \ref{conv} is a straightforward corollary of
Condition C1 and the fact that the functions $F_n(z)$ are
equicontinuous on any compact subset of $\CP$.
If $q$ is unbounded then one needs to show additionally that
the contribution of the tails of $dN_n(E)$ to $F_n(z)$
is negligible in the limit
$n\to\infty$. Obviously, it will suffice to prove that:
\begin{equation}\label{pb}
\hbox{for any $\varepsilon >0$ there is a $B>0$ such that
$\frac{1}{n} \sum\limits_{k=1}^n \chi_B(|E_k|)  \ln (1+|E_k|) <
\varepsilon$ for all n},
\end{equation}
where the summation in (\ref{pb}) is effectively over all
eigenvalues of $H_n^0$ such that $|E_j|\ge B$. To complete the
proof note that, in view of the inequalities in (\ref{ll}), it is
apparent that Condition C2* implies (\ref{pb}). \hfill $\Box$

\medskip

We finish this Section with a proof of the log-H\"older continuity
of $N(E)$. This property is well known for random potentials and
in this case it follows from the fact that $U(x,y)\ge 0$
\cite{CS}. In turn, this inequality is a consequence of the
Thouless formula according to which $U(x,y)$ coincides with the
Lyapunov exponent of $H_n^0$ with $n=\infty$, see e.g.
\cite{CS,PF,CL}.

In our case Conditions C1 and C2 are too weak to guarantee the
existence of the Lyapunov exponent even for non-real values of the
spectral parameter, see examples in Appendix \ref{appendix2}.
However these two conditions ensure that the function $U(x,y)$ is
bounded from below which, in turn, implies (very much in the same
ways as in \cite{CS}) that $N(E)$ is log-H\"older continuous.

\medskip

\begin{Thm} \label{sec1:thm1} Assume C1-C2. Then:
\begin{itemize}
\item[(i)] $U(x,y)\ge -c_0$ for some $c_0>0$ and all real $x$ and $y$.
\item[(ii)] $N(E)$ is log-H\"older continuous: for any $E$ and
$|\sigma|\leq\frac{1}{2}$
\begin{equation}\label{1c}
|N(E+\sigma)-N(E)|= c(E,\sigma)|\ln|\sigma||^{-1}\ \hbox{ where }\
\lim_{\sigma\to 0}c(E,\sigma)=0.
\end{equation}
If $E$ belongs to a compact set then $c(E,\sigma)\leq c_1$ with
the constant $c_1$ depending only on this compact set and the
constant $C$ in Condition C2.
\end{itemize}
\end{Thm}
\emph{Proof.} Part (i).  In view of (\ref{i4}) and the symmetry in
$y$, it will suffice to prove the inequality for $y>0$ only. It
follows from Theorem \ref{sec1:thm2} that
\begin{equation}\label{i5}
\liminf_{n\to\infty}U_n(x,y)=U(x,y)+c_0
\end{equation}
for some $c_0\ge 0$ and all $x$ and $y\not=0$. To finish the
proof, it is sufficient to show that the LHS in (\ref{i5}) is
non-negative. If the $\lim_{n\to\infty} U_n(x,y)$ exists then it
coincides with the Lyapunov exponent, and hence is non-negative.
The general case can be treated similarly (see Appendix
\ref{appendix1}).

Part (ii). Since $\int \ln |x-E| dN(E) \ge -c_0$, we have that
\[
\int_{|x-E|\le 1}\ln |x-E|\, dN(E) + \int_{|x-E|>1}\ln |x-E|\,
dN(E) \ge -c_0,
\]
and
\[
\int\limits_{|x-E|\le 1}\ln \frac{1}{|x-E|} \, dN(E) \le
\int\limits_{|x-E|>1}\ln |x-E| \, dN(E)  +c_0 \le U(x,1)+c_0.
\]
Therefore, for any $|\delta| \le \frac{1}{2}$,
\[
U(x,1)+c_0 \ge \int\limits_{|x-E|\le |\delta| }\ln \frac{1}{|x-E|}
\, dN(E) \ge |N(x+\delta)-N(x-\delta)|\ln \frac{1}{|\delta|},
\]
and
\begin{equation}\label{cont}
|N(x+\delta)-N(x-\delta)| \le [U(x,1)+c_0] |\ln |\delta||^{-1}.
\end{equation}
Note that for any compact set $K\subset \R$, $\max_K U(x,1)<
+\infty$. This is because $U(x,1)$ is continuous in $x$.

Now, define for  $|\delta| \le \frac{1}{2}$
\[
c (x,\delta) = \int_{x}^{x+\delta} \ln \frac{1}{|x-E|}\, dN(E).
\]
Obviously,
\[
|N(x+\delta)-N(x)| \le \frac{c(x,\delta)}{|\ln |\delta||}.
\]
To complete the proof, note that (\ref{cont}) implies that the
measure $dN(E)$ has no atoms, and therefore
\[
c (x,\delta) \longrightarrow 0 \hspace{2ex} \hbox{when $\delta \to
0 $.}
\]
\hfill $\Box$

\section{Main results}
\label{sec3}

For the sake of convenience and clarity of exposition, we shall
formulate and prove our results for the class of potentials $q$
satisfying Conditions C1 and C2*. We emphasize however that our
main results hold true, modulo trivial modifications, under
Conditions C1 and C2, and the corresponding proofs are identical
to those given in Section \ref{sec4}. This is a mere reflection of
Theorem \ref{sec1:thm2} and the fact that our proofs are based on
the convergence of $F_n(z)$ to $F(z)$.

\subsection{Notations and auxiliary statements}

Let us fix any finite interval $[a,b]$ of the real axis $(a<b)$.
Most of our results apply to the part of the spectrum of $H_n^g$
belonging to the strip $\{z:\ \re z\in[a,b]\}$ in the complex
plane $\C$.

We define several critical values of the parameter $g$:
\begin{equation}\label{crit}
\underline{\hspace{0.5ex}g}_{\mathrm{cr}}=\inf_{x\in \mathcal{S}}
U(x,0), \hspace{3ex} \bar{g}_{\mathrm{cr}}=\sup_{x\in \mathcal{S}}
U(x,0).
\end{equation}
where we have introduced the notation $\mathcal{S}$ for the
support of the measure $dN(E)$, and
\begin{equation}\label{crit1}
\underline{\hspace{0.5ex}g}_{\mathrm{cr}}(a,b)=\inf_{x\in
\mathcal{S}\cap[a,b]} U(x,0), \hspace{3ex}
\bar{g}_{\mathrm{cr}}(a,b)=\sup_{x\in \mathcal{S}\cap[a,b]}
U(x,0).
\end{equation}
It may happen that $\bar {g}_{\mathrm{cr}}=+\infty$, and it is
obvious, in view of Theorem \ref{sec1:thm1}, that
$\underline{\hspace{0.5ex}g}_{\mathrm{cr}} \ge 0$ for any
potential satisfying Conditions C1 and C2.

\medskip

For every $g\in \R$, define
\begin{equation}\label{lg}
\Lambda_g=\{x\in \R: \hspace{1ex} U(x,0 )< g \}.
\end{equation}
If $g\le \underline{\hspace{0.5ex}g}_{\mathrm{cr}}$ then
$\Lambda_g=\emptyset$, otherwise $\Lambda_g$ consists of (possibly
infinitely many) disjoint open intervals:
\begin{equation}\label{ab}
\Lambda_g=\bigcup_j \, (a_j,b_j).
\end{equation}
We note also that $U(x,y)=U(x,-y)$ and that
\begin{equation}\label{pos}
\frac{\partial}{\partial y}\, U (x,y) >0 \hspace{2ex} \hbox{for
any} \hspace{1ex} x\in \R, \hspace{1ex} y > 0.
\end{equation}

\medskip

\begin{Prop} \label{sec0:prop1} Suppose that $g>
\underline{\hspace{0.5ex}g}_{\mathrm{cr}}$
and let  $(a_j,b_j)$ be the intervals defined in (\ref{ab}). Then
the level set ${\cal L}_g=\{(x,y): \hspace{1ex} U(x,y)=g,
\hspace{1ex} y>0 \}$ consists of disjoint analytic arcs
\begin{equation}\label{arcs}
y=y_j(x), \hspace{2ex} a_j<x <b_j,
\end{equation}
whose end-points lie on the real axis, i.e.
$y_j(a_j+0)=y_j(b_j-0)=0$, if $-\infty < a_j, b_j< +\infty$.
\end{Prop}
\emph{Proof}. If $x_0\notin \Lambda_g$ then $U(x_0,y) > g$ for all
$y\not=0$. Therefore the equation $U(x_0,y)=g$ cannot be solved
for $y>0$.

Consider now any of the intervals $(a_j,b_j)$. If $x_0\in
(a_j,b_j)$ then $U(x_0,0)<g$, and in view of (\ref{pos}) and
$U(x,+\infty)=+\infty$, there exists a unique positive solution
$y_0\overset{\rm def}{=}  y_j(x_0)>0$ of the equation
$U(x_0,y)=g$. As $U(x,y)$ can be
analytically continued into a neighborhood of $(x_0,y_0)$ in
$\C^2$, the implicit function theorem asserts that $y_j(x)$ is
analytic in a disk $|x-x_0| < \delta$ in the complex $x$-plane.
The union of all such disks, when $x_0$ runs through $(a_j,b_j)$
covers $(a_j,b_j)$. Therefore the function $y_j(x)$ can be
analytically continued into a domain in the complex $x$-plane that
contains $(a_j,b_j)$, and, for any closed interval
$[\alpha,\beta]\subset (a_j,b_j)$, this domain contains
\begin{equation}\label{D}
D_{\alpha,\beta}= \{x\in \C: \hspace{1ex} \alpha-h \le \re x \le
\beta+h, \hspace{1ex} |\im x| \le h \}
\end{equation}
for some $h>0$.

If  $a_j>-\infty$ then $y_j(a_j+0)=0$. For, if not then
$\bar{y}:=\limsup_{x\to a_j, x>a_j} y_j(x)>0$. But then
$U(a_j,\bar{y})=g$ and hence $U(x,0)<g$ for every $x$ from some
neighborhood of $a_j$ which contradicts the definition of $a_j$ as
the end point of our interval.
The same argument proves that if $b_j< +\infty$ then
$y_j(b_j-0)=0$. \hfill $\Box$

\subsection{Statement of results}

We are now in a position to formulate our main results.

\medskip

\begin{Thm} \label{sec2:thm0}
For any $g > 0$ all the eigenvalues of $H_n^g$ belong to the level
lines of the function $U_n(x,y)$ defined by the equation
\begin{equation}\label{2c}
U_n(x,y)=g + \frac{2}{n}\ln (1- e^{-ng}).
\end{equation}
\end{Thm}

\medskip

\begin{Thm} \label{sec2:thm1} (i) Suppose that $g\leq
\underline{\hspace{0.5ex}g}_{\mathrm{cr}}(a,b)$. Then for any
$\varepsilon>0$ there exists $n_0= n_0(\varepsilon,g,q,a,b)$ such
that for any $n>n_0$ all the eigenvalues $z_j$ of $H_n^g$ with
$\re z_j\in[a,b]$ belong to the $\varepsilon$-neighborhood of the
real axis: $|\im z_j| \leq \varepsilon$.

(ii) Suppose that  $g > \underline{\hspace{0.5ex}g}_{\mathrm{cr}}$
and $(a_j,b_j)$ is one of the intervals comprising $\Lambda_g$.
Then for any $[\alpha,\beta]\subset (a_j,b_j)$ there exists $n_1=
n_1(q,g,\alpha,\beta)$ such that for any $n>n_1$ there exists a
solution $y_{j,n}(x)$ to equation (\ref{2c}) which is analytic in
the domain $D_{\alpha,\beta}$ defined in (\ref{D}) and
\begin{equation}\label{2d}
\lim_{n\to\infty}y_{j,n}(x)=y_{j}(x), \hspace{3ex} \hbox{uniformly
in $x \in D_{\alpha,\beta}$}.
\end{equation}
The function
$y_{j,n}(x)$, for $n>n_1$, is the only solution of (\ref{2c})
which is non-negative when $x\in [\alpha,\beta]$.
\end{Thm}

\emph{Remarks.} 1.
The previous two theorems imply that if $H_n^g$
has eigenvalues in the strip
$a_j< \alpha\leq\re z\leq\beta<b_j$, then, for $n>n_1$,
they must lie on the analytic arc
\[
{\cal A}_n(\alpha,\beta)=\{(x,y):\hspace{1ex} y=y_{j,n}(x),
\hspace{0.5ex} \alpha \le x \le \beta \}
\]
and on its reflection with respect to the real axis.

2. Relation (\ref{2d}) implies that the arcs ${\cal
A}_n(\alpha,\beta)$ converge to the level lines of
$U(x,y)$ when $n\to\infty$ together with all their derivatives.

3. We did not make use of two out of the four critical values introduced 
in (\ref{crit}) and (\ref{crit1}). However, their role is clear: if 
$g> \bar{g}_{\mathrm{cr}}(a,b)$ then no limiting eigenvalue curves 
grow out of the support of $ \mathcal{S}\cap[a,b]$. 
If $\bar{g}_{\mathrm{cr}}$ is finite then $[a,b]$
can be replaced in this statement by $ \mathcal{S}$.

\smallskip
The next two theorems describe the asymptotic distribution of the
eigenvalues of $H_n^g$ along the arc ${\cal A}_n(\alpha,\beta)$.
In particular they state that $H_n^g$, for large $n$,
does have eigenvalues on ${\cal A}_n(\alpha,\beta)$.

By $\nu_n(\alpha ,\beta)$ we denote  the number
of complex eigenvalues of
$H_n^g$ lying on ${\cal A}_n(\alpha,\beta)$.

\medskip

\begin{Thm} \label{sec2:thm2}
For any closed interval $[\alpha,\beta] \subset (a_j,b_j)$,
\[
\lim_{n\to\infty} \frac{1}{n}\nu_n(\alpha , \beta) =\frac{1}{2\pi}
[\theta (\beta)- \theta (\alpha)],
\]
where $\theta(x)=-V(x,y_j(x))$ and $V(x,y)$ is the imaginary part
of $F(z)$.
\end{Thm}

\medskip

\emph{Remark.} Let $l$ be the natural parameter on the curve
$y=y_j(x)$, that is the length of the part of this curve contained
between say $(\alpha,y_j(\alpha))$ and $(x,y_j(x))$. A simple
calculation involving the Cauchy-Riemann equations for $F(z)$
shows that $d\theta=|f(z(l)|dl$, where
\begin{equation}\label{f(z)}
f(z)=F^{'}(z)=\int\frac{dN(E)}{z-E}.
\end{equation}
Hence
\begin{equation}
\lim_{n\to\infty} \frac{1}{n}\nu_n(\alpha, \beta) =
\frac{1}{2\pi}\int_{\alpha}^{\beta} \theta'(x)
dx=\frac{1}{2\pi}\int_{\alpha+iy_j(\alpha)}^{\beta
+iy_j(\beta)} |f(z(l))| dl,
\end{equation}
where the integration is carried out along the path $y=y_j(x)$
from $\alpha+iy_j(\alpha)$ to $\beta+iy_j(\beta)$.

Theorems \ref{sec2:thm1} and \ref{sec2:thm2} are not entirely new
and can be inferred from Theorems 2.1 and 2.2 in \cite{GK2}. We
are now going to formulate our principal new result. Let $[\alpha,
\beta]$ be the same as before. Let us label the eigenvalues
$z_k=x_k+iy_k$ of $H_n^g$ lying on the arc ${\cal
A}_n(\alpha,\beta)$ so that $\alpha\leq x_1\leq x_2\ldots \leq
x_m\leq\beta$ (we note that in fact the multiplicity of these
eigenvalues is one and the inequalities here are strict; this
follows from the inequality $\theta_n^{'}(x)\ge C_0 >0$  which is
a part of the proof of Theorem  \ref{sec2:thm2}).

\medskip

\begin{Thm} \label{sec2:thm4} For any two consecutive eigenvalues
$z_k$ and $z_{k+1}$ of $H_n^g$ lying on ${\cal A}_n (\alpha,
\beta)$,
\begin{equation} \label{z}
n(z_{k+1}-z_k)=\frac{2\pi i}{f(z_k)} + \delta_n(z_k, z_{k+1})
\end{equation}
where
\begin{equation} \label{z1}
\lim_{n\to\infty}\delta_n(z_k,
z_{k+1}) = 0 \hspace{3ex} \hbox{uniformly in} \hspace{1ex} z_k,
z_{k+1} \in {\cal A}_n(\alpha,\beta).
\end{equation}
\end{Thm}

\section{Proofs}\label{sec4}

The eigenvalues and eigenfunctions of $H_n^g$ are determined by
the equation
\begin{equation}\label{main1}
-e^g\varphi_{k+1}+q_k\varphi_k-e^{-g}\varphi_{k-1}=z\varphi_k,\ \
1\leq k\leq n
\end{equation}
were
\begin{equation}\label{bc1}
\varphi_0=\varphi_n,\ \varphi_1=\varphi_{n+1}.
\end{equation}
The parameter $g$ can be eliminated from (\ref{main1}) by making
use of the standard substitution $\varphi_k=e^{-kg}\psi_k$ which
transforms (\ref{main1}) into
\begin{equation}\label{main2}
-\psi_{k+1}+q_k\psi_k-\psi_{k-1}=z\psi_k,\ \ 1\leq k\leq n,
\end{equation}
and boundary conditions (\ref{bc1}) into
\begin{equation}\label{bc2}
\psi_0=e^{-ng}\psi_n,\ \psi_1=e^{-ng}\psi_{n+1}.
\end{equation}
Note that the transformed boundary conditions are asymmetric
(unless $g=0$).

To solve equation (\ref{main2}) we shall follow the standard routine and
rewrite it in the matrix form:
\[
 \left(
\begin{array}{l}
\psi_{k+1} \\ \psi_k \\
\end{array}
\right) =A_k \left(
\begin{array}{l}
\psi_{k} \\ \psi_{k-1} \\
\end{array}
\right),\ 0\leq k \leq  n, \hspace{1ex} \hbox{ where }\ A_k=
\begin{pmatrix}
q_{k}-z  & -1    \\1   &  0    \\
\end{pmatrix}.
\]
Then
\[
\begin{pmatrix}
\psi_{n+1} \\ \psi_n \\
\end{pmatrix}
=S_n(z)
\begin{pmatrix}
\psi_{1} \\ \psi_{0} \\
\end{pmatrix}
,\ \hbox{ where }\ S_n(z)=A_nA_{n-1}\ldots A_1.
\]
On the other hand,
\[
\begin{pmatrix}
\psi_{n+1} \\ \psi_n \\
\end{pmatrix}
=e^{ng}
\begin{pmatrix}
\psi_{1} \\ \psi_{0} \\
\end{pmatrix}
\]
because of boundary conditions (\ref{bc2}). Therefore the
eigenvalues of $H_n^g$ are determined by the equation
\begin{equation}\label{33}
\det[S_n(z)-e^{ng}I]=0.
\end{equation}
Since $\det S_n(z)=1$ for all $z$, we have that
\[
\det[S_n(z)-e^{ng}I] = 1- e^{ng}\tr S_n (z)+ e^{2ng}.
\]
Hence:

\medskip

\begin{Lemma}\label{lemma1} $z$ is an eigenvalue of
$H_n^g$ iff
\[
\tr S_n (z)=e^{ng}+e^{-ng}.
\]
\end{Lemma}

\medskip

The trace of the matrix $S_n(z)$ is a polynomial in $z$ of degree
$n$. The following representation of this polynomial, which is
well known in the context of the discrete Hill equation (see e.g.
\cite{T} or \cite{L}) is useful for our purposes.

\medskip

\begin{Lemma}\label{lemma2}
Let $E_j$, $j=1,\ldots, n$,  be the eigenvalues of $H_n^0$. Then
\begin{equation}\label{det1}
\tr S_n(z) = \prod_{j=1}^{n}(E_j-z)+2.
\end{equation}
\end{Lemma}
\emph{Proof}. For $g=0$, Lemma \ref{lemma1} asserts that the
polynomials $\prod_{j=1}^{n}(E_j-z)$ and $\tr S_n(z)-2$ have the
same set of zeros. It is easy to verify both polynomials have the
same coefficient, $(-1)^n$,  in front of the highest power of $z$,
hence they must coincide. \hfill $\Box$

\medskip

Here is our main technical lemma:

\medskip

\begin{Lemma}\label{lemma3} Suppose that $g>0$. Then $z$ is
an eigenvalue of $H_n^g$ iff
\begin{equation}\label{2b}
F_n(z)=\frac{2}{n} \ln \Big(e^\frac{ng}{2}-
e^\frac{-ng}{2}\Big)+\frac{i\pi}{n} \hspace{5ex}
\Big(\hspace{-2ex}\mod \frac{2\pi i}{n}\Big)
\end{equation}
where $F_n(z)$ is the function defined in (\ref{2}).
\end{Lemma}
\emph{Proof}. It follows from Lemmas \ref{lemma1} and \ref{lemma2}
that $z$ is an eigenvalue of $H_n^g$ iff
\begin{equation}\label{final}
\prod_{j=1}^{n}(E_j-z)=- (e^\frac{ng}{2}- e^\frac{-ng}{2})^2.
\end{equation}
Since $F_n(z)=\frac{1}{n}\sum_{j=1}^n \ln (E_j-z)$, equation
(\ref{final}) is equivalent to (\ref{2b}), provided $g\not=0$.
\hfill $\Box$

\medskip

We are now in a position to prove Theorems \ref{sec2:thm0} and
\ref{sec2:thm1}.

\medskip

\noindent \underline{{Proof of Theorem \ref{sec2:thm0}}}: This
theorem is a straightforward corollary of Lemma \ref{lemma3}.

\medskip

\noindent \underline{{Proof of Theorem \ref{sec2:thm1}}}: (i) Let
$g(\varepsilon)=\min_{x\in [a,b]} U(x,\varepsilon)$. According to
Proposition \ref{conv} one can find $n_0$ such that for all $x\in
[a,b]$
\[
|U_n(x,\varepsilon)-U(x,\varepsilon)| \le
\frac{1}{2}[g(\varepsilon) -
\underline{\hspace{0.5ex}g}_{\mathrm{cr}}(a,b)]
\]
if $n>n_0$. (Note that $g(\varepsilon)>
\underline{\hspace{0.5ex}g}_{\mathrm{cr}}(a,b)$ because of
(\ref{pos}).) Thus, for all $a\le x \le b$ and $y\ge \varepsilon$,
\[
U_n(x,y) \ge U_n(x,\varepsilon)\ge
U(x,\varepsilon)-\frac{1}{2}[g(\varepsilon) -
\underline{\hspace{0.5ex}g}_{\mathrm{cr}}(a,b)] \ge \frac{1}{2}
U(x,\varepsilon)+\frac{1}{2}\underline{\hspace{0.5ex}g}_{\mathrm{cr}}(a,b)
\ge \underline{\hspace{0.5ex}g}_{\mathrm{cr}}(a,b).
\]
Recall that by the assumption
$\underline{\hspace{0.5ex}g}_{\mathrm{cr}}(a,b) \ge g$. Since $g>
g + \frac{2}{n} \ln (1-e^{-ng})$ for any $n>0$, we conclude that,
for all $n>n_0$ equation (\ref{2c}) has no solutions in the
half-strip $a \le  x \le b$, $y\ge \varepsilon$. To complete the
proof remember that $U_n(x,-y)=U_n(x,y)$.

(ii) First, note that for every real $x$ equation (\ref{2c}) has
one non-negative solution at most.

Now, let $g>\underline{\hspace{0.5ex}g}_{\mathrm{cr}}$ and
$[\alpha,\beta]\subset (a_j,b_j)$ where $(a_j,b_j)$ is one of the
intervals comprising $\Lambda_g$. Recall that $y_j(x)$ is analytic
in $D_{\alpha,\beta}$, see Proposition \ref{sec0:prop1}. Because
of the compactness of $[\alpha,\beta]$, it will suffice to prove
the existence of the solution $y_{j,n}(x)$ to equation (\ref{2c})
and its convergence to $y_j(x)$ as $n\to\infty$ in a
small neighborhood of every point $(x,y_j(x))$ where $x$ runs
through $[\alpha,\beta]$.

Fix $\tilde x \in [\alpha,\beta]$ and consider the point $(\tilde
x, \tilde y)$ where $\tilde y=y_j(\tilde x)$. It follows from the
integral representations for $U_n(x,y)$ and $U(x,y)$ that these
two functions are analytic in the domain
\[
\tilde D\overset{\rm def}{=}
\{(x,y): \hspace{1ex} |x-\tilde x| < \frac{\tilde y}{2},
\hspace{0.5ex} |y-\tilde y| < \frac{\tilde y}{2} \}.
\]
We shall use the following general lemma. Put $D_r\overset{\rm def}{=}
\{(x,y): \hspace{1ex} |x-\tilde x| < r,
\hspace{0.5ex} |y-\tilde y| < r \}$.

\medskip
\begin{Lemma}\label{lemma4}
Let $\Phi (x,y)$ and $\tilde \Phi(x,y)$ be two functions analytic
in $D_r$ and such that for all $(x,y)\in D_r$
\[
|\Phi_{x}^{'}(x,y)|\le c_1, \hspace{1ex} 0<c_2\le |
\Phi_{y}^{'}(x,y)|\le c_3, \hspace{1ex} |\tilde \Phi (x,y)|\le 1.
\]
Suppose that $\Phi (\tilde x,\tilde y)=0$. Then there is a
positive $\varepsilon_0$ which depends only on $c_1,c_2,c_3,$ and $r$ (but
not on $\Phi(\cdot,\cdot)$, $\tilde \Phi(\cdot,\cdot)$!) such that
the equation
\begin{equation}\label{ift}
\Phi (x,y)+\varepsilon \tilde \Phi(x,y)=0
\end{equation}
has a unique solution $y=y(x,\varepsilon)$ which is analytic in
$(x,\varepsilon)$ in the domain $\{(x,\varepsilon): \hspace{1ex} |x-\tilde x|<2 \varepsilon_0, \,
|\varepsilon|<2 \varepsilon_0 \}$ and $y(\tilde x,0)=\tilde y$.
\end{Lemma}
\emph{Proof of Lemma \ref{lemma4}}. Consider the function $G
(x,\varepsilon,y)\overset{\rm def}{=}\Phi (x,y)+\varepsilon \tilde
\Phi(x,y)$ of three complex variables $x$, $\varepsilon$, and $y$.
In the domain $D_{\frac{r}{2}}$ we have:
\[
\left|\tilde\Phi_x^{'}(x,y)\right|=(2\pi)^{-1}\left|\int_{|u|=r}
\Phi(u,y)(x-u)^{-2}du\right|\leq{\frac{2}{r}}
\] and similarly $\left|\Phi_y^{'}\right|\leq{\frac{2}{r}}$. Hence, for
$\varepsilon$ sufficiently small, $G_x^{'}$ and $G_y^{'}$ are close to
$\Phi_x^{'}$ and $\Phi_y^{'}$ correspondingly. It is clear that
$\left|G_{\varepsilon}^{'}\right|\leq 1$.
The implicit function theorem for an analytic function of three variables $x$,
$\varepsilon$, and $y$ implies now the existence of the
solution $y=y(x,\varepsilon)$ to the equation $G (x,\varepsilon,y)=0$.
 It should be emphasized that the domain where this solution exists
and is analytic depends only on the corresponding estimates of $G_x^{'}$,
$G_y^{'}$, and $G_{\varepsilon}^{'}$. $\Box$

\medskip

To finish our proof of Theorem \ref{sec2:thm1}, note that in our
case $U(x,y)$ plays the role of $\Phi (x,y)$ and equation
(\ref{ift}) has the form
\[
U(x,y)+\varepsilon \frac{U_n(x,y)-U(x,y)}{\varepsilon_0}=0.
\]
Here we first choose $\varepsilon_{0}$ so that to satisfy
the conditions of Lemma \ref{lemma4}, and then choose $n_0$
such that $|U_n(x,y)-U(x,y)|\le \varepsilon_0$ for all
$(x,y)\in \tilde D$ and $n>n_0$. The wanted result follows from our Lemma
when $\varepsilon=\varepsilon_0$.
\hfill $\Box$

\medskip
Define
\[
\theta_n(x)=-V_n(x,y_{j,n}(x)) \hspace{2ex} \hbox{and}
\hspace{2ex} \theta (x)=-V(x,y_{j}(x))
\]
for $x\in [\alpha,\beta]\subset (a_j,b_j)$ and $n>n_1$ with
$n_1$ as in part (ii) of  Theorem \ref{sec2:thm1}. As before,
$V_n(x,y)$ and $V(x,y)$ are the imaginary parts of the analytic
functions $F_n(z$ and $F(z)$, see (\ref{3}) and (\ref{2}). In view
of Theorem \ref{sec2:thm1} and Proposition \ref{sec0:prop1},
we have that
\begin{equation}\label{t}
\lim_{n\to \infty}\theta_{n}(x) = \theta (x) \hspace{3ex}
\hbox{uniformly in $x\in [\alpha, \beta ]$}.
\end{equation}
It follows from the Cauchy-Riemann equations for $F_n(z)$ and
$F(z)$ that
\begin{equation}\label{tdash1}
\theta_n^{'}(x)=\left. \frac{|\nabla
U_n(x,y)|^2}{\frac{\partial}{\partial y}U_n(x,y)}
\right|_{y=y_{j,n}(x)} \hspace{2ex} \hbox{and} \hspace{2ex}
\theta^{'}(x)=\left. \frac{|\nabla
U(x,y)|^2}{\frac{\partial}{\partial y}U(x,y)} \right|_{y=y_{j}(x)}.
\end{equation}
Therefore we also have that
\begin{equation}\label{tdash}
\lim_{n\to \infty}\theta_{n}^{'}(x) = \theta^{'}(x) \hspace{3ex}
\hbox{uniformly in $x\in [\alpha, \beta]$}.
\end{equation}
As $\frac{\partial}{\partial y}U_n(x,y)$ and
$\frac{\partial}{\partial y}U(x,y)$ are positive in the upper half
of the $xy$-plane, the functions $\theta_n(x)$ and $\theta(x)$
are
monotone increasing. Moreover, in view of (\ref{tdash}), it
apparent that there is a constant $C_0>0$ such that
\begin{equation}\label{pos3}
\theta_n^{'}(x)\ge C_0 >0 \hspace{3ex}
\hbox{for every $x\in [\alpha,\beta]$ and $n>n_1$}.
\end{equation}
Now we are in a position to prove Theorems \ref{sec2:thm2} and
\ref{sec2:thm4}.

\medskip

\noindent \underline{{Proof of Theorem \ref{sec2:thm2}}}: On the
arc ${\cal A}_n(\alpha,\beta)$, i.e. when $z=x+iy_{j,n}(x)$,
$\alpha \le x \le \beta$, the eigenvalue equation (\ref{2b})
reduces to
\begin{equation}\label{c}
e^{in\theta_n(x)}=-1.
\end{equation}
When $x$ runs through $[\alpha,\beta]$ in the positive direction,
$\theta_n(x)$ gets positive increment and $w=e^{in\theta_n(x)}$
moves anticlockwise along the unit circle $|w|=1$.
Obviously, $\nu_n(\alpha,\beta)$, the number of
eigenvalues of $H_n^g$ on ${\cal A}_n(\alpha,\beta)$, is, up to
$\pm 1$, equal to the number of circuits completed by $w$ when $x$
completes its run. Thus
\[
\nu_n(\alpha,\beta)= \lfloor\; \frac{n[\theta_{n}(\alpha)-
\theta_{n}(\beta)]}{2\pi}\; \rfloor +\kappa,
\]
where $|\kappa| \le 1$.
When $n\to\infty$, $\theta_n(x)$ converges to $\theta
(x)=-V(x,y_j(x))$, and, therefore, $\frac{1}{n}
\nu_n(\alpha,\beta)$ converges to  $\frac{1}{2\pi} [\theta
(\beta)-\theta (\alpha) ]$.  \hfill $\Box$

\medskip

\noindent\underline{{Proof of Theorem \ref{sec2:thm4}}}: Let
$z_l=x_l+iy_j(x_l)$ and $z_{l+1}=x_{l+1}+iy_j(x_{l+1})$ be two
consecutive eigenvalues of $H_n^g$ on ${\cal
A}_{n}(\alpha,\beta)$. We assume that $x_{l+1} > x_l$. It follows
from equation (\ref{c}) that
\[
\theta_n(x_{l+1})-\theta_n(x_{l})=\frac{2\pi}{n},
\]
and therefore
\begin{equation}\label{x3}
x_{l+1} - x_l=\frac{2\pi}{n}\frac{1}{\theta^{'}_n(x^*)}
\end{equation}
for some $x^*\in (x_l,x_{l+1})$. In view of (\ref{pos3}),
\[
0 < x_{l+1}-x_l \le \frac{2\pi}{C_0}\frac{1}{n}.
\]
for all $n$ large. Hence $x^{*} \to x_l$ as $n\to\infty$, and
(\ref{x3}) and (\ref{tdash}) imply that
\begin{equation}\label{x}
n(x_{l+1}-x_l)=\frac{2\pi}{\theta^{'}(x_l)}+\delta_{n}(x_l,x_{l+1})
\end{equation}
where
\begin{equation}\label{x1}
\lim_{n\to\infty}\delta_{n}(x_l,x_{l+1})=0 \hspace{3ex}
\hbox{uniformly in $x_l,x_{l+1}\in
[\alpha,\beta]$}.
\end{equation}

To prove (\ref{z}) -- (\ref{z1}), note that
\[
z_{l+1}-z_l=x_{l+1}-x_l +iy_{j,n}^{'}(x^{**})(x_{l+1}-x_l)
\]
for some $x^{**} \in (x_l, x_{l+1})$.
By making use of (\ref{x}), one
obtains that
\[
n(z_{l+1}-z_l)=\frac{2\pi}{\theta (x_l)}
[1+iy_{j,n}^{'}(x^{**})]+\delta_n(x_l,x_{l+1}).
\]
Now (\ref{z}) -- (\ref{z1}) easily follow from Theorem
\ref{sec2:thm1} and the following relation
\begin{equation}\label{h}
\frac{1+iy_j^{'}(x)}{\theta^{'}(x)}=\left. \frac{1}{i
F^{'}(z)}\right|_{z=x+iy_j(x)}.
\end{equation}
To verify this relation, make use of the equation
\[
\int \log |x+iy_j(x)-E|dN(E)=g,
\]
to obtain
\[
y_j^{'}(x)=-\left. \frac{U^{'}_x(x,y)}{U^{'}_y(x,y)
}\right|_{y=y_j(x)}.
\]
Now, in view of (\ref{tdash1}),
\[
\frac{1+iy_j^{'}(x)}{\theta^{'}(x)}=\left.
\frac{1}{U^{'}_y(x,y)+iU^{'}_x(x,y)}\right|_{y=y_j(x)}.
\]
and (\ref{h}) follows by the Cauchy-Riemann equations for $F(z)$.
\hfill $\Box$

\appendix

\section{Appendix} \label{appendix1}

\begin{Prop} \label{prop:a1}
For all real $x$ and $y\not=0$ we have
\begin{equation}\label{a0}
\liminf_{n\to\infty}U_n(x,y)\ge 0
\end{equation}
\end{Prop}
\emph{Proof}. This result can be proved in many ways. We present
here a proof based on (\ref{det1}).

It follows from (\ref{det1})
that
\begin{equation}\label{a2}
\frac{1}{n}\ln \tr S_n(z) - F_n(z) =\frac{1}{n}\ln
\left[1+\frac{2}{\prod_{j=1}^n (E_j-z)}\right]
\end{equation}
Therefore for every $z\in\CP$ such that $\im z > 1$
\begin{equation}\label{a1}
\lim_{n\to\infty} \left[\frac{1}{n}\ln \tr S_n(z) -
F_n(z)\right]=0.
\end{equation}
The two functions in the LHS in (\ref{a2}) are analytic and
uniformly bounded in $n$ on compact sets in $\CP$. Therefore, by
the Vitali theorem, (\ref{a1}) must hold for every $z\in \CP$.

Consider now the eigenvalue equation for $S_n(z)$. If $z\notin \R$
then $S_n(z)$ has no eigenvalues on the unit circle. As $\det
S_n(z)=1$, we then have that for every non-real $z$
the $2\times 2$ matrix $S_n(z)$ has one eigenvalue, $\lambda_n(z)$, in the
exterior of the unit circle, i.e. $|\lambda_n(z)| >1$, and the
other one, $1/\lambda_n(z)$, in the interior of the unit circle.
Thus
$
\tr S_n(z)=\lambda_n(z)+\lambda^{-1}_n(z)
$
and
\begin{equation}\label{a3}
\frac{1}{n}\ln \lambda_n(z) -\frac{1}{n}\ln \tr S_n(z)=\frac{1}{n}
\ln [1+\lambda^{-2}(z)].
\end{equation}
It follows from (\ref{det1}) that $|\tr S_n(z)|$ grows
exponentially fast with $n$ provided $|\im z|>1$, and then so does
the dominant eigenvalue of $S_n(z)$. This is because
$
|\tr S_n(z)| \le |\lambda_n(z)|+1
$
(In fact, $\lambda_n(z)$ grows exponentially fast with $n$ for
every non-real $z$.) Hence, for every $z\in \CP$ such that $\im
z>1$,
\begin{equation}\label{a4}
\lim_{n\to\infty} \left[\frac{1}{n} \ln \lambda_n(z) -
\frac{1}{n}\ln \tr S_n(z)\right]=0.
\end{equation}
The two functions in the LHS in (\ref{a3}) are analytic and
uniformly bounded in $n$ on compact sets in $\CP$. Therefore, by
the Vitali theorem, (\ref{a4}) holds for every $z\in \CP$. From
(\ref{a1}) and (\ref{a4}) we have that
\[
\lim_{n\to\infty} \left[\frac{1}{n} \ln \lambda_n(z)
-F_n(z)\right]=0
\]
for  every $z\in \CP$. Taking the real part,
\[
\lim_{n\to\infty} \left[\frac{1}{n} \ln |\lambda_n(x+iy)|
-U_n(x,y)\right]=0,
\]
and therefore (recall that $|\lambda_n(z)| \ge 1$)
\[
\liminf_{n\to\infty} U_n(x,y) =\liminf_{n\to\infty}\frac{1}{n} \ln
|\lambda_n(x+iy)|\ge 0.
\]
\hfill $\Box$

\section{Appendix}\label{appendix2}

Obviously the integrated density of states, $N(E)$, depends on the
potential $q$. To make this dependence explicit, we shall write in this
section $N_n(E;q)$ and $N(E;q)$ instead of $N_n(E)$ and $N(E)$.

Let $k_1, k_2, \ldots $ be an increasing (infinite)
sequence of natural numbers such that
\begin{equation}\label{i9}
\frac{\#\{j: \hspace{0.5ex} k_j \le n\}}{n} \longrightarrow 0 \hspace{3ex}
\hbox{as $n\to\infty$},
\end{equation}
and let $v=\{v_k\}_{k=1}^{\infty}$ be a potential supported by the
sequence $k_j$, i.e. $v_k=0$ unless $k\in \{k_j\}$.

\medskip
\begin{Prop}\label{b:prop1} If $q$ is a selfaveraging potential then so is
$\tilde q=q+v$, and $N(E; \tilde q)=N(E,q)$.
\end{Prop}
\emph{Proof.}
According to the well known theorem from linear algebra, if
$A$ and $B$ are two selfadjoint $n\times n$ matrices then the number of
eigenvalues of the matrix $A+B$ in interval $\Delta$ differs
from that of the matrix $A$ by rank$(A-B)$ at most.  Hence
\[
|N_n(E;\tilde q)-N_n(E;q)|\le \frac{\#\{j: \hspace{0.5ex} k_j \le n\}}{n},
\]
which proves the proposition. \hfill $\Box$

\medskip

Let us now assume that the potential $v$ satisfies Condition C2
and $|v_{k_j}|\to \infty$ as $j\to\infty$. Define
\[
s_n=\frac{1}{n}\sum_{j:\, k_j\le n} \ln (1+|v_{k_j}|).
\]
and
\[
\tilde U_n(x,y)=\int_{-\infty}^{+\infty} \ln |x+iy-E| dN(E; \tilde
q), \hspace{3ex} y>0.
\]

\medskip

\begin{Thm} \label{b:thm1}
Let $q$ be a selfaveraging potential with $\sup_k |q_k|=M$, and
$\tilde q=q+v$. Then
\[
\lim_{n\to\infty} [\tilde U_{n}(x,y)-s_n] = U(x,y).
\]
\end{Thm}
\emph{Proof.} We prove this Theorem under the following additional
condition on $v$: $|v_{k_{j+1}}|
> |v_{k_{j}}|+2+M$ for all $j$. The proof for the
general case requires minor but cumbersome modifications.

Let $z=x+iy$. By definition, 
\[
\tilde U_{n}(x,y)=\frac{1}{n}\sum_{k=1}^n \ln |z-\tilde E_k| =
\frac{1}{n}\sum_{|\tilde E_k|\le 2+M} \ln |z-\tilde E_k| +
\frac{1}{n}\sum_{|\tilde E_k| > 2+M} \ln |z-\tilde E_k|,
\]
where the $\tilde E_k$ are the eigenvalues of $H_n^0$ with the
potential $\tilde q$. By Proposition \ref{b:prop1} the first sum
converges to $U(x,y)$ when $n\to\infty$. We note next that the
eigenvalues $\tilde E_k$ in the interval $|\tilde E_k| > 2+M$ have
the following property: for all but may be a finite number of them
there is a unique $j$ such that $|\tilde E_k - v_{k_j}| \le 2+M$.
This is due to the fact that the eigenvalues of the operator of
multiplication by $v$ differ from the eigenvalues of $H_n^0$ with
the potential $\tilde q$ by $2+M$ at most. Hence taking into
account that $\ln \frac{|z-\tilde E_k|}{1+|v_{k_j}|} \to 0$ when
$|E_k|\to\infty$, we obtain that
\[
\lim_{n\to \infty}
\left[
\frac{1}{n} \sum_{|\tilde E_k| > 2+M} \ln |z-\tilde E_k| -
\frac{1}{n} \sum_{k_j \le n, \, |v_{k_j}| > 2+M} \ln (1+|v_{k_j}|)
\right]=0.
\]
It is apparent that
\[
\lim_{n\to\infty} \left[\frac{1}{n} \sum_{k_j \le n, \,
 |v_{k_j}| > 2+M} \ln (1+|v_{k_j}|) -s_n\right]=0,
\]
which completes the proof. \hfill $\Box$

\medskip

It is easy now to construct examples showing that the statement
of Proposition \ref{prop:liminf} cannot be improved.

\medskip

\emph{Example 1.} Let $k_j=j^2$ and $v_{j^2}=e^{j}$, $j=1,2,
\ldots .$ Then $\lim_{n\to\infty}s_n = \frac{1}{2}$. Hence
$\lim_{n\to\infty} \tilde U_n(x,y)=U(x,y)+\frac{1}{2}$.

\medskip

\emph{Example 2.} Let $k_j=2^j$ and $v_{2^j}=e^{2^{j}}$, $j=1,2,
\ldots .$ Then $\limsup_{n\to\infty} s_n =2$  and
$\liminf_{n\to\infty} s_n=1$. Hence $\limsup_{n\to\infty} \tilde
U_n(x,y)=U(x,y)+2$ and $\liminf_{n\to\infty} \tilde
U_n(x,y)=U(x,y)+1$.  It is easy to check that for every
\marginpar{c!} $1\le c\le 2$ there is a subsequence $U_{n_j}(x,y)$
converging to $U(x,y)+c$ when $j\to\infty$.

\end{document}